\documentclass[10pt,letterpaper,twocolumn]{article} 

\usepackage{ol2}
\usepackage[draft]{hyperref}
\usepackage{amsmath}

\begin{document}

\twocolumn[ 
\hyphenation{pro-ba-bi-li-ty}
\title{Super-bunched bright squeezed vacuum state}


\author{T.~Sh.~Iskhakov$^{1,*}$, A.~Perez$^1$, K.~Yu.~Spasibko$^2$, M.~V.~Chekhova$^{1,2}$, G.~Leuchs$^1$}

\address{
$^1$Max-Planck Institute for the Science of Light, \\  Guenther-Scharowsky-Str. 1 / Bau 24, Erlangen  D-91058, Germany \\
$^2$Physics Department, Moscow State University, \\ Leninskiye Gory 1-2, Moscow 119991, Russia
\\

$^*$Corresponding author: Timur.Iskhakov@mpl.mpg.de
}

\begin{abstract} In this paper we experimentally study the statistical properties of a bright squeezed vacuum state containing up to $10^{13}$ photons per mode ($10~\mu J$ per pulse), produced via high gain parametric down conversion (PDC). The effects of bunching and superbunching of photons were observed for a single mode PDC radiation by second-order intensity correlation function measurements with analog detectors.

\end{abstract}


 ] 

\noindent Parametric down conversion (PDC) is one of the widespread sources of nonclassical light fields. There are two main classes of states that can be obtained via PDC in experiment. First, few-photon states: for example, conditionally prepared Fock states\cite{Lvovsky2001}, two-photon Bell states\cite{Burlakov2002}, polarization-entangled W and GHZ states~\cite{Kobayashi2004}. These states have found their applications in quantum information and quantum computing. Second, macroscopic quantum states: quadrature\cite{Ling-An} and two-mode squeezed states~\cite{Aytur}. The noise of these states is reduced below the shot noise level and therefore they are attractive for high-precision measurements and optical communication. Another type of a macroscopic quantum state, intensively studied at present, is bright squeezed vacuum (BSV). The quantum properties of this state are mostly of interest in the single-mode case. However, since the BSV generated via PDC in a bulk crystal is usually multi-mode with typical coherence time of few tenths of a picosecond, consequently the observation of single-mode statistics of BSV is a difficult experimental problem.

It is well known that the statistics of a single mode of signal or idler squeezed vacuum radiation is thermal~\cite{Glauber}. It was experimentally proved by Tapster and Rarity~\cite{Tapster}, who observed the statistics of one spatiotemporal mode of PDC signal beam by measuring the second-order intensity correlation function (ICF) in a standard Hanbury Brown-Twiss interferometer. The obtained value of correlation function was $g^{(2)}=1.85$, while the expected value for the thermal light is $g^{(2)}_{th}=2$. The single-mode statistics was also observed by narrow spectral and spatial filtering of the signal PDC beam in~\cite{Gizin2004} and for nearly single-mode states generated in a periodically poled $KTiOPO_4$ waveguide~\cite{Silberhorn2011}.  Multi-mode thermal statistics of the signal beam of PDC radiation was measured in~\cite{Bondani2004} and~\cite{Hubel2009}. In both experiments the number of detected modes was about twenty so the observed statistics was almost Poissonian. It should be noted here that the single-mode states obtained earlier always contained very low photon numbers and therefore they cannot be used for the conditional preparation of large photon-number Fock states.

In this letter we present an experiment dedicated to the observation of the statistics of an extremely bright single-mode squeezed vacuum state. Special attention is paid to the detection conditions to be satisfied in order to observe the statistics of one mode of PDC radiation. We consider two states generated under two different phase matching conditions: collinear nondegenerate and collinear degenerate. The effects of  photon bunching and superbunching were experimentally observed for these states, respectively.

The size of a spatiotemporal mode of the radiation is approximately given by the coherence volume~\cite{Klyshkobook},

\begin{equation}
V_{coh} \simeq c\cdot\tau_{coh}\cdot \rho_{coh}^2,
\label{volume}
\end{equation}
where $c$ is the phase velocity of light in a medium, $\tau_{coh}$ is the coherence time, and $\rho_{coh}^2$ is the coherence area. Inside the coherence volume the degree of coherence is not uniform: it is the highest at the center and decreases closer to the borders because of the influence of the fields that belong to the neighboring coherence volumes. Therefore the single-mode statistics can be observed if the detection volume $V_{det}$ is much smaller than the coherence volume,
\begin{equation}
V_{det}\equiv c\cdot\tau_{det}\cdot\rho_{det}^2 \ll V_{coh},
\label{ID}
\end{equation}
where $\tau_{det}$ and $\rho_{det}^2$ denote the detection time and the detection area, respectively.

The twin-beam squeezed vacuum state at the output of a nondegenerate collinear optical parametrical amplifier (NOPA) manifests perfect photon number $n$ correlation in the conjugate channels, $|\psi\rangle=\sum\limits_{n=0}^\infty{C_n|n,n\rangle}$. Here the coefficients $C_n$ depend on the parametric gain. The photon number distribution in each channel is `geometric',  $P_{B}(n)=\frac{\langle n\rangle^n}{{(\langle n\rangle+1)}^{n+1}}$, with the mean photon number $\langle n\rangle$. Under the condition (\ref{ID}), photon bunching can be observed via second-order ICF measurement: $g^{(2)}=2$.

The single-mode squeezed vacuum state at the output of a degenerate OPA contains only even photon numbers: $|\psi\rangle=\sum\limits_{n=0}^\infty{C_n|2n\rangle}$. The photon number probability distribution~\cite{Perina}, $P_{SB}(n)=\frac{n!}{2^{n}(\frac{n}{2}!)^2}\frac{\langle n\rangle^{n/2}}{(\langle n\rangle+1)^{n/2+1/2}}$ for even $n$ and $P_{SB}(n)=0$ for odd $n$, allows us to observe the effect of superbunching $(g^{(2)}>2)$ in the Hanbury Brown-Twiss measurement~\cite{Janszky,Loudon},
\begin{equation}
g^{(2)}=3+\frac{1}{N},
\label{SB}
\end{equation}
where $N$ is the mean photon number per mode, i.e., in the coherence volume of PDC radiation. In the case of detecting $m$ modes $(m=V_{det}/V_{coh}\gg1)$ the value of measured ICF is reduced by a factor of $m$ in comparison with the single-mode case:
\begin{equation}
g_{meas}^{(2)}=1+\frac{g^{(2)}-1}{m}.
\label{mm}
\end{equation}
Therefore multimode detection makes the superbunching effect for spontaneous PDC only observable when $m\cdot N<1$. Taking into account that the number of photons per mode increases as $N=\sinh^2\Gamma$, where $\Gamma$ is the parametric gain, at high gain the second term in (\ref{SB}) is eliminated and the effect of superbunching can be observed only at low number of detected modes. The superbunching at high-gain PDC was already observed by means of an ultrafast measurement, namely, two-photon counting in semiconductors~\cite{Boitier}. However, sometimes single-mode BSV is required for further applications. In this case, a single mode of BSV should be filtered out optically.


In order to generate states manifesting photon bunching and superbunching we have built the experimental setup presented in Fig. 1. The BSV state was produced in two 2 mm $\mathrm{BBO}$ crystals by pumping with the pulsed radiation of the third harmonic of a Nd:YAG laser at 355 nm, with a pulse duration of 18 ps and a repetition rate of 1 kHz. The pump power was varied by means of a half wave plate ($HWP$) and a Glan prism ($GP$) placed in front of the crystals. The pump was focused by a 100 cm lens ($L_p$) into a waist of $0.26$ mm diameter to increase the parametric gain. The effect of spatial walk off in two crystals was reduced twofold by placing the optical axes of the crystals in the horizontal plane but in opposite directions. The crystals were cut for type-I collinear degenerate phase matching, the central wavelength being $\lambda_{c}=709.3\,\hbox{nm}$ and the PDC spectral width (FWHM) was $57$ nm. After eliminating the UV pump by a cut-off mirror ($M$) and a long-pass filter ($OG$), the angular spectrum of parametric radiation was selected by an aperture ($A$) placed in the focal plane of a 30 cm lens ($L$). The spectral mode filtering was made by a monochromator with a spectral resolution of $\Delta\lambda=0.1\hbox{nm}$. To increase the signal the light was focused by a lens ($L_M$) on the input slit of a monochromator. At the output of the monochromator the radiation was split at a nonpolarizing beamsplitter ($NPBS$) into two channels and focused by two lenses ($L_1$, $L_2$) on a pair of analog photodetectors ($D_1$, $D_2$) based on {\itshape{p-i-n}} diodes in combination with low-noise amplifiers described in~\cite{Iskhakov2009}. The signals from the detectors were measured by an analog-digital converter by integrating the electronic pulses over time. The mean values of the signals per pulse $\langle S_1\rangle$, $\langle S_2\rangle$ and the mean of the product signals $\langle S_1\cdot S_2\rangle$ were used for the calculation of the ICF: $g^{(2)}_{meas}=\frac{\langle S_1\cdot S_2\rangle}{\langle S_1\rangle \cdot \langle S_2\rangle}$. Before the main experiment the setup was calibrated by measuring the second-order ICF for coherent radiation. As a source we used a strongly attenuated second harmonic of the Nd:YAG laser at 532 nm. The obtained result $g^{(2)}_{coh}=1.005\pm0.005$ is in agreement with the expected value and therefore confirms the validity of the measurement procedure.

\begin{figure} [h]
\centerline{\includegraphics[width=8cm]{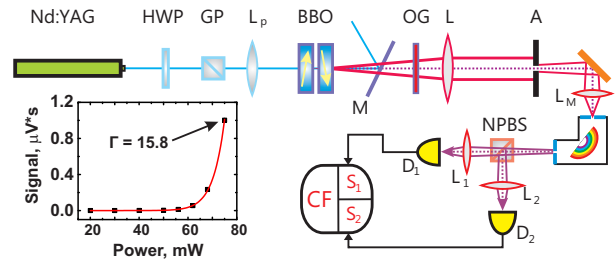}}
\caption{The experimental setup. Bottom left: Dependence of the PDC signal on the pump power.}
\label{fig:1}
\end{figure}

First, the parametric gain of collinear degenerate PDC was determined through the nonlinear dependence of the PDC signal on the power of the pump (the inset of Fig. 1). The highest value achieved in the measurement with $75$ mW mean pump power was $ \Gamma_{max}=15.8$, which corresponds to extremely high mean number of photons per mode: $N\approx1.3\cdot 10^{13}$.

\begin{figure}
[h]\centerline{\includegraphics[width=9.0cm]{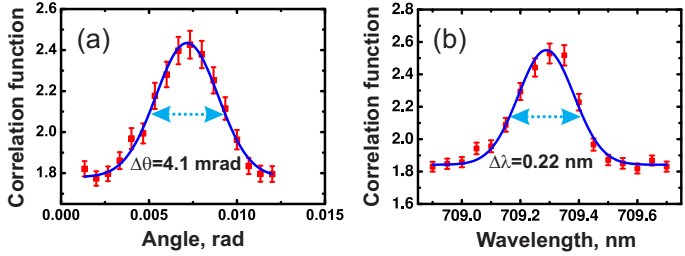}}
\caption{The second-order ICF measured versus the angle $(a)$ and the wavelength (b).}
\label{fig:2}
\end{figure}

The angular (spatial) correlation function for degenerate PDC was measured by scanning the aperture ($A$) in the horizontal direction in the far-field plane (Fig. \ref{fig:2} (a)). In order to select the degenerate wavelength of PDC radiation the monochormator was adjusted to transmit $\lambda_c=(709.3\pm0.05)$ nm. As predicted by the theory, the central part of the dependence represents the effect of photon superbunching, when the pinhole selects the signal and the idler beams simultaneously. The ICF values at the edges of this dependence represents the effect of photon bunching when only signal or idler beam is detected. The blue line is a Gaussian fit to determine the FWHM of the correlation area (the spatial mode size), which is $(4.1\pm 0.3)$ mrad.

Figure \ref{fig:2} (b) shows the spectral distribution of $g^{(2)}$ measured for collinear PDC. Similar to the angular distribution the data reveal the superbunching of photons for the degenerate case when $\lambda_i=\lambda_s$ and the bunching of photons for nondegenerate PDC. The FWHM of this dependence ($0.22\pm0.03$ nm) provides the information about the spectral width of the correlation.


\begin{figure}
[h]\centerline{\includegraphics[width=5.0cm]{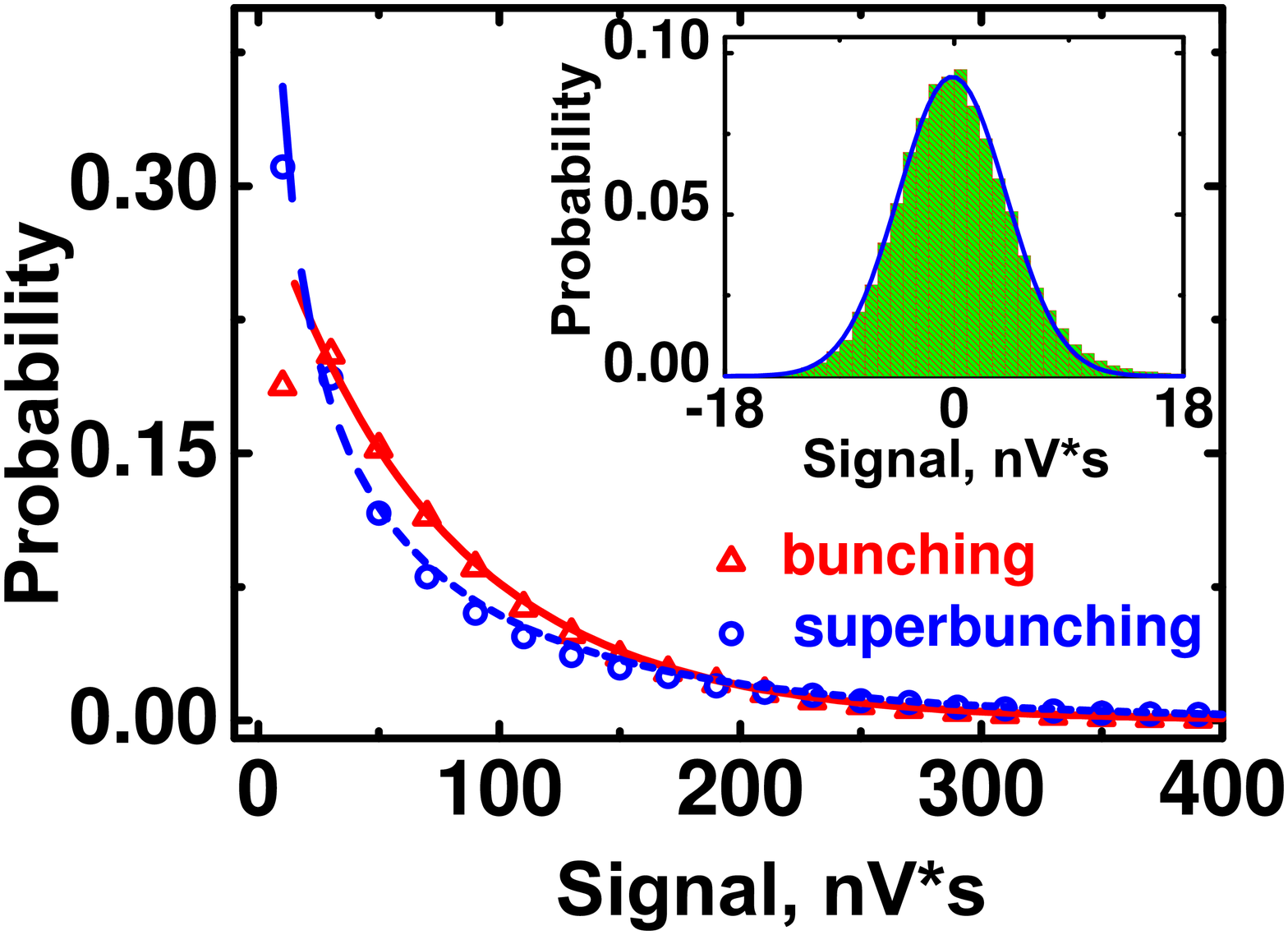}}
\caption{Signal probability distribution for thermal (red triangles) and superthermal (blue circles) statistics. The curves are the theoretical fits. Inset: probability distribution for the electronic noise.}
\label{fig:3}
\end{figure}

The measured signal probability distributions for the radiation at $708$ nm
(red triangles) and $709.3$ nm (blue circles) are presented in Figure \ref{fig:3} . The solid red curve and the dashed blue one are the theoretical fits plotted according to the probability distributions $P_{B}(n)$ and $P_{SB}(n)$ respectively with the fixed mean signals. One can see that the distributions are different, indicating that different regimes of the PDC have different photon-number statistics. The losses in the optical channel smooth out only the internal structure of the probability distribution for the state with even photon numbers, however, the envelope does not change. The mean values of the signals are about $70$ nV$\cdot$s, which corresponds to $8\cdot10^{3}$ detected photons~\cite{Iskhakov2009}. Both probability distributions are influenced around zero by the electronic noise of the detectors. Nevertheless this influence is not dramatic: the $FWHM=10$ nV$\cdot$s of the electronic noise histogram (see the inset of Fig. \ref{fig:3}) is almost by an order of magnitude smaller than the mean values of the signals.


Finally, the question is why the measured values of the ICF are less than expected from the theory, i.e., 3 for the degenerate case and 2 for the nondegenerate case. According to (\ref{ID}) the single mode statistics can be observed when the detection volume is much smaller then the coherence volume. In this experiment the detected angle $\theta_{det}=0.45$ mrad was indeed much smaller then the angular mode size $\theta= 4.1$ mrad. Meanwhile the best spectral bandpass of the monochromator $\Delta\lambda_{det}=0.1$ nm was just twice as narrow as the spectral mode size $\Delta\lambda_{mode}=0.22$ nm. Thereby the condition (\ref{ID}) was not completely satisfied and the effective number of detected modes, according to (\ref{mm}), was $m = 1.25$.

Therefore the state of light after the angular and frequency filtering is nearly single-mode.

In conclusion, we have achieved extremely high parametric gain (15.8) in a single-pass OPA. Nearly single-mode selection $(m=1.25)$ was carried out. The photon bunching was observed for collinear non-degenerate PDC. For collinear degenerate PDC, superbunching of photons was observed. The states produced in the experiment were so bright that even after single-mode selection the mean photon number was about $8 000$ photons per pulse.

This work was supported in part by the Russian Foundation for Basic Research, grant \# 10-02-00202 and 11-02-01074. T.~Sh.~I. acknowledges support from Alexander von Humboldt Foundation.

\end{document}